# Ferroelastic-switching-driven colossal shear strain and piezoelectricity in a hybrid ferroelectric

Yuzhong Hu[1,2,†], Lu You[3,†], Bin Xu[3,†], Tao Li[2], Samuel Alexander Morris[5], Yongxin Li[6], Yehui Zhang[7], Xin Wang[7], Pooi See Lee[2], Hong Jin Fan[1,*] and Junling Wang[2,4,*]

[1]School of Physical and Mathematical Sciences, Nanyang Technological University, Singapore 637371

[2]School of Materials Science and Engineering, Nanyang Technological University, Singapore 639798

[3]Jiangsu Key Laboratory of Thin Films, School of Physical Science and Technology, Soochow University, 1 Shizi Street, Suzhou 215006, China

[4]Department of Physics, Southern University of Science and Technology, Shenzhen 518055, China

[5]Facility for Analysis, Characterisation, Testing and Simulation (FACTS), Nanyang Technological University, Singapore, 639798

[6]Division of Chemistry and Biological Chemistry, School of Physical and Mathematical Sciences, Nanyang Technological University, Singapore 637371

[7]School of Sciences, Nantong University, Nantong 226019, China

†These authors contributed equally to this work.

*Corresponding author. Email: jlwang@ntu.edu.sg or jwang@sustech.edu.cn (J.W.); fanhj@ntu.edu.sg (H.J.F.)

**Materials that can produce large controllable strains are widely used in shape memory devices[1], actuators[2,3] and sensors[4]. Great efforts have been made to improve the strain outputs of various material systems. Among them, ferroelastic transitions underpin giant reversible strains in electrically-driven ferro/piezoelectrics and thermally- or magnetically-driven shape memory alloys. However, large-strain ferroelastic switching in conventional ferroelectrics is very challenging while magnetic and thermal controls are not desirable for applications. Here, we demonstrate an unprecedentedly large shear strain up to 21.5 % in a hybrid ferroelectric, $C_6H_5N(CH_3)_3CdCl_3$. The strain response is about two orders of magnitude higher than those of top-performing conventional ferroelectric polymers and oxides. It is achieved via inorganic bond switching and facilitated by the structural confinement of the large organic moieties, which prevents the undesired 180° polarization switching. Furthermore, Br substitution can effectively soften the bonds and result in giant shear piezoelectric coefficient ($d_{35}$ ~ 4800 pm/V) in Br-rich end of the solid solution, $C_6H_5N(CH_3)_3CdBr_{3x}Cl_{3(1-x)}$. The superior electromechanical properties of the compounds promise their potential in lightweight and high energy density devices, and the strategy described here should inspire the development of next-generation piezoelectrics and electroactive materials based on hybrid ferroelectrics.**

The archetype of shape memory alloys, NiTi, can produce a strain of around 8 % from a temperature induced reversible phase transition[5]. Other shape memory alloys can be triggered by magnetic fields. NiMnGa, for instance, shows up to 10 % shear strain due to twin boundary movements under a magnetic field[4]. Perovskite oxide ferroelectrics can achieve a strain of 0.3 %[6]. Through point-defect engineering and ferroelastic domain switching, their strain response can be increased up to 0.75 %[7]. For polymer systems, various stimuli have been exploited to generate strains, including change in pH, temperature, light, moisture, magnetic and electric fields[8]. Due to their structural flexibility, the strain response of some polymers, e.g., elastomers, can even surpass 100 %[9].

However, although many materials are available for various applications, very few satisfy the requirements of large strain, high energy density and high speed simultaneously. For example, shape memory alloys can produce large stress and medium strain, but the actuation output is always at a low speed[10]. Piezoelectric oxides can generate large stress at high speed, but the strain response is much lower than that of other materials. Polymers, on the contrary, produces the largest strain but one to two orders of magnitude lower actuation stress (2 MPa on average)[11]. Organic-inorganic hybrid ferroelectrics (OIHF) comprise of inorganic frameworks filled with organic moieties. It is thus possible that they can take advantages of both the lightweight, flexible organic moiety and the mechanically strong and tough inorganic backbone, producing large strain with high energy/power output[12,13]. In recent years, various OIHFs have been designed and investigated, particularly, making great progress towards piezoelectric and electrostrictive applications[14-16]. Piezoelectric responses of some OIHFs have even outperformed traditional oxide ferroelectrics such as $BaTiO_3$ and $Pb(Zr, Ti)O_3$ (PZT) [17,18]. The multiaxial nature of these OIHFs implies that ferroelastic switching may lie at the core of the enhanced piezoelectricity[19]. However, direct proof of deterministic control of ferroelastic switching is still lacking.

The ferroelectricity in most OIHFs originates from the off-center ordering of the positively charged organic moieties, which link with the inorganic frameworks through hydrogen bond and van der Waals interaction. The reversal of the spontaneous polarization sometimes requires a complete flip of the large organic moiety, in stark contrast to oxide ferroelectrics, where polarization switching only involves small displacement of the ions. It is thus possible to suppress the strain-equivalent 180° polarization switching in OIHFs with large molecular moiety due to the high energy barrier and promote ferroelastic switching with large strain response. Following this strategy, we synthesize and investigate a hybrid ferroelectric, $C_6H_5N(CH_3)_3CdCl_3$ ((PTMA)CdCl$_3$), which produces colossal ferroelastic shear strain about two orders of magnitude higher than that of lead-based piezoelectrics and larger than all shape memory alloys. The non-volatile, reversible strain is attributed to macroscopic ferroelastic switching reflecting atomic-level lattice distortions, thanks to the structural confinement effect. Furthermore, we demonstrate that Br substitution effectively softens the bonds and flattens the switching energy landscape, leading to giant shear piezoelectric response in Br-rich compounds of the solid solution (PTMA)CdBr$_{3x}$Cl$_{3(1-x)}$.

X-ray diffraction (XRD) and differential scanning calorimetry (DSC) measurements were carried out to investigate the crystal structures and phase transitions of (PTMA)CdBr$_{3x}$Cl$_{3(1-x)}$. Going from Cl to Br, the lattice constants of the compound increase continuously (see Table 1 and Extended Data Fig. 1), while phase transition temperature keeps decreasing from above 180 °C (decomposition temperature) to 19 °C (see Extended Data Fig. 2). However, no structural transition is induced by Br substitution. Powder XRD refinements indicate that the solid solutions all crystallize in monoclinic *Cc* space group in the low temperature phase (LTP) and orthorhombic *Ama2* in the high temperature phase (HTP) (Table 1 and Extended Data Table 1), which can be described by the Aizu notation of *mm2*F*m* among the 88 ferroelectric transition species[20]. In the LTP, (PTMA)CdBr$_{3x}$Cl$_{3(1-x)}$ consists of PTMA organic moieties and a scaffold of 1D edge-sharing cadmium-halide (CdX$_5^-$) hexahedra (see Fig. 1 and Extended Data Fig. 3 for pure Cl, Br and mixed halide structures, respectively). They are connected via hydrogen bonds between the methyl group and the halide anions. In the HTP, the longest Cd-X bond of the hexahedron breaks, transforming into corner-sharing CdX$_4^-$ tetrahedron. In the meantime, order-disorder phase transition is triggered in the organic part with two degenerate positions and the restoration of mirror symmetry with respect to the (001) plane.

Table 1| Structural information of (PTMA)CdBr$_{3x}$Cl$_{3(1-x)}$ in the LTP.

| Br amount (%) | 0 | 15 | 29 | 45 | 63 | 70 | 77 | 80 | 90 |
|---|---|---|---|---|---|---|---|---|---|
| a / Å | 12.7541 | 12.8262 | 12.8497 | 12.8960 | 12.9338 | 12.9763 | 12.9909 | 13.0015 | 13.0465 |
| b / Å | 14.5035 | 14.5961 | 14.6257 | 14.6785 | 14.7126 | 14.7458 | 14.7443 | 14.7601 | 14.7827 |
| c / Å | 7.1515 | 7.2275 | 7.2554 | 7.31681 | 7.3650 | 7.4153 | 7.4261 | 7.4567 | 7.5058 |
| β / ° | 96.3069 | 95.910 | 95.823 | 95.5064 | 95.282 | 94.985 | 95.035 | 95.000 | 94.841 |
| Vol / Å³ | 1314.86 | 1345.89 | 1356.50 | 1378.63 | 1395.54 | 1413.54 | 1416.92 | 1425.52 | 1442.43 |
| Space Group | *Cc* | *Cc* | *Cc* | *Cc* | *Cc* | *Cc* | *Cc* | *Cc* | *Cc* |

The ferroelectric polarization of (PTMA)CdBr$_{3x}$Cl$_{3(1-x)}$ in the LTP can be understood by regarding N and Cd sites as the centers of positive and negative charges in the unit cell, respectively. As shown in Fig. 1b, the spontaneous polarization has components along both the *a* and *c* axes. However, due to the large size of the organic moiety and confined space between the inorganic chains, a 180° reversal of the polarization is unlikely. This is evidenced by the fact that the polarization along the *a* axis persists even in the HTP. Fortunately, polarization switching along the *c* axis is not hindered by the structure, which will rotate the polarization vector and generate large shear strain as shown in Fig. 1b. Polarization-electric field (*P-E*) hysteresis loops can only be obtained along the **c** axis, while no polarization switching is observed along the *a* axis prior to electrical breakdown. The unique polarization switching path is consistent with the *mm*2F*m* Aizu notion, suggesting that symmetry breaking across the phase transition occurs only in one axis. Remarkably, upon polarization switching, the macroscopic tilting angle of the (PTMA)CdCl$_3$ crystal exhibits an appreciable change (~ 12.8 °) that perfectly correlates with the unit cell distortion ($\Delta\beta = 12.6$ °) (Fig. 1c, also see Supplementary Movies S1 and S2), which translates to a shear strain of ~ 22 % ($\tan\Delta\beta$).

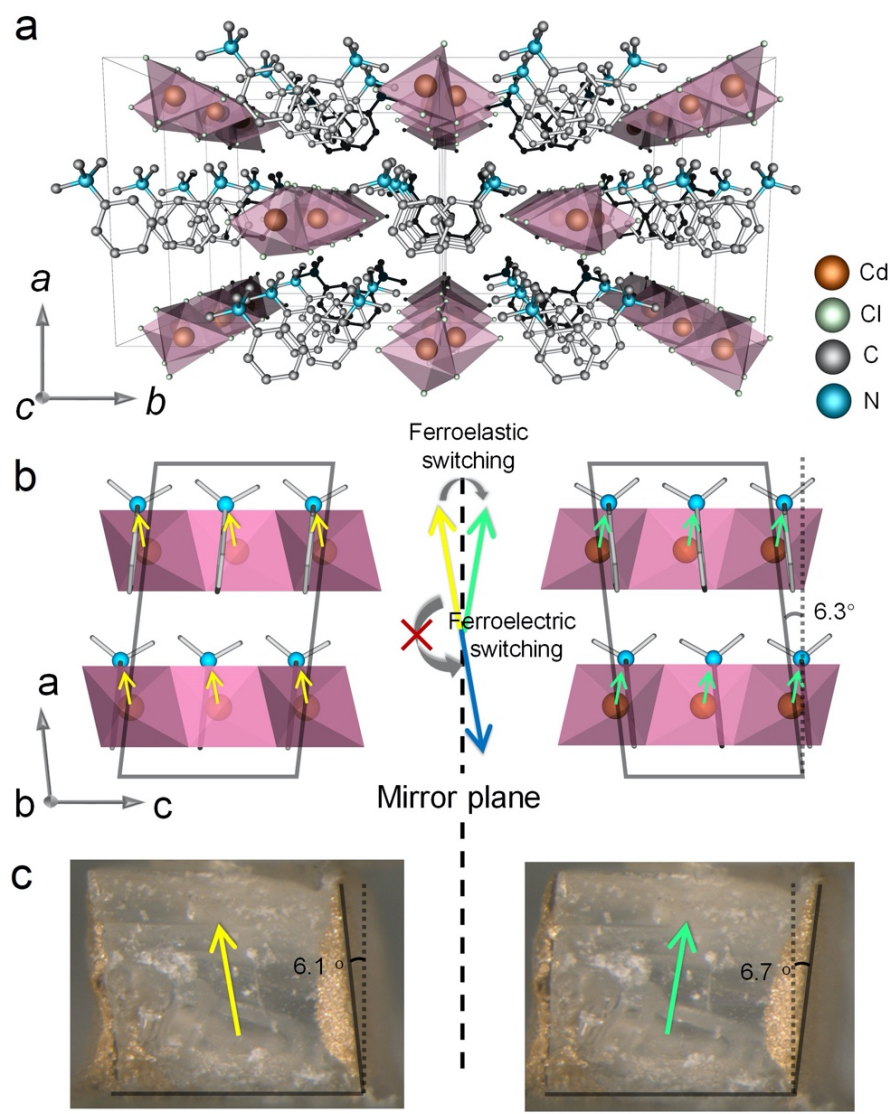

**Fig. 1 | Crystal structures and pictures of (PTMA)CdCl$_3$ at two different polarization/strain states. a**, Perspective of crystallographic structure along the 1D chain direction. **b**, Structural units at the two polarization states and structure confinement effect for ferroelectric/ferroelastic switching. Yellow and green arrows indicate polarization directions of each organic moiety-metal halide anion unit. **c**, Pictures of a crystal at two polarization states, showing the macroscopic shear strain.

Quantitative polarization and shear strain measurements were performed on the solid solutions by simultaneously recording the *P-E* and strain-electric field (*S-E*) hysteresis loops. Note that polarization of this material series cannot be switched below a critical temperature because of the large coercivity compared to breakdown field, so the measurements were conducted at elevated temperatures. As shown in Fig. 2a, with increasing Br content, the remnant polarization ($P_r$) decreases monotonically from 3.6 µC/cm² to 3.0 µC/cm², which agrees well with the calculated polarizations of 3.7 µC/cm² for pure Cl compound and 2.9 µC/cm² for the Br counterpart. Additionally, Br substitution apparently reduces the energy barrier for polarization switching, as indicated by the reduced critical temperature and coercivity. Shear strain (*S*) in this case is defined as $S = d/H = \tan\Delta\beta$ [21], where *d* is the shear strain induced movement of the upper surface, *H* is the height of sample and $\Delta\beta$ is the shear angle (Extended Data Fig. 4). The reversible shear strain hits a record-breaking value of ~ 21.5 % in the pure Cl compound, and decreases with increasing Br content, which agrees with the smaller *β* angle in Br-rich compounds (Table 1). The Young's modulus of (PTMA)CdCl₃ is about 5.0 GPa based on our first principles calculations. In Fig. 2c, we compare reported actuation strain and volumetric energy density for various material systems (see references in Extended Data Table 2). The maximum shear strain of (PTMA)CdCl₃ is about two orders of magnitude higher than those of conventional ferroelectrics, surpassing all the reported shape memory alloys. Note that electroactive polymers (EAPs) produce strains based on Maxwell stress, and the electric field needed are in the order of MV/cm, much larger than the field used here. And for temperature-driven shape memory alloys (TSMAs) and shape memory polymers (SMPs), the response time is very long. So the appropriate comparison should be with ferroelectric oxides (FE oxides), piezoelectric polymers and magnetically-driven shape memory alloys (MSMAs), and (PTMA)CdCl₃ performs better than all of them.

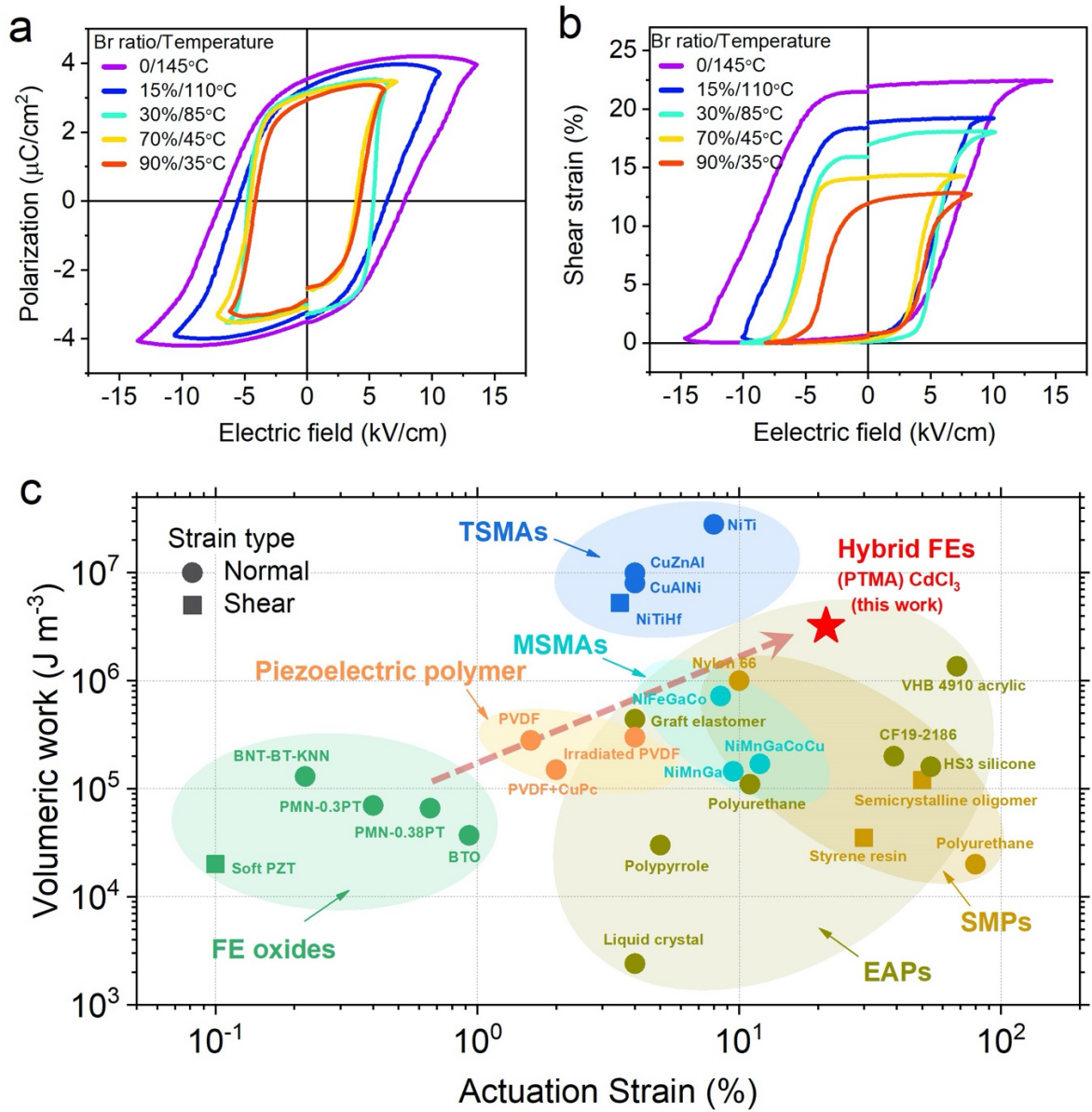

**Fig. 2 | Ferroelectric and ferroelastic properties of (PTMA)CdBr$_{3x}$Cl$_{3(1-x)}$. a**, *P-E* loops and **b**, shear *S-E* loops of (PTMA)CdBr$_{3x}$Cl$_{3(1-x)}$ for selected compositions at a measurement frequency of 2Hz. **c**, Actuation strain and work outputs for materials of different systems, i.e., ferroelectric (FE) oxides, piezoelectric polymer, magnetic-driven shape memory alloys (MSMAs), temperature-driven shape memory alloys (TSMAs), electroactive polymers (EAPs), shape memory polymers (SMPs) and hybrid FE (see Extended Data Table 2 for references).

More intriguingly, with Br substitution, the lattice is softened significantly and giant shear piezoelectric response is achieved in Br-rich compounds. Unipolar piezoelectric response can be measured on samples that have been pre-poled using the same experimental setup shown in Extended Data Fig. 4. Using Voigt notation, the piezoelectric coefficient measured in this case is $d_{35}$, which can be calculated by $d_{35} = S_5/E_3$, namely, the slope of the saturation region of the $S$ – $E$ curves. As shown in Fig. 3a, $d_{35}$ increases monotonically with more Br substitution and reaches an average value of around 4800 pm/V for 90% Br compound. The giant shear piezoelectric coefficient is higher than recently discovered organic[23], hybrid[18] piezoelectrics as well as other classical piezoelectrics based on BaTiO$_3$ and PbTiO$_3$ (Fig. 3d)[24,25]. The greatly enhanced piezoelectric response in Br-rich crystals is attributed to the chemical-substitution induced bond softening. This not only increases the intrinsic electrostrictive response of the compound, but also facilitates ferroelastic (back)switching after the sample has been pre-poled. This is evidenced in Fig. 3b, c, in which the piezoelectric behavior of 90% Br sample shows obvious field and frequency dependence, a signature of nonlinear elastic domain contributions. Under large field (along the same direction as the pre-poling field) at low frequency, the back switched ferroelastic domains are activated again and contribute to the observed piezoelectric response. However, the involvement of ferroelastic switching inevitably leads to hysteresis effect, which could be alleviated via targeted chemical engineering to achieve relaxor-like properties[26,27]. This softening effect can also be seen in composition-dependent *P-E* loop measurements (Fig. 2a) and dielectric study. The field and frequency dependent dielectric response shows that dielectric constant increases much more obviously for 90% Br compound than that of the pure Cl one under high field at low frequency (Extended Data Fig. 5).

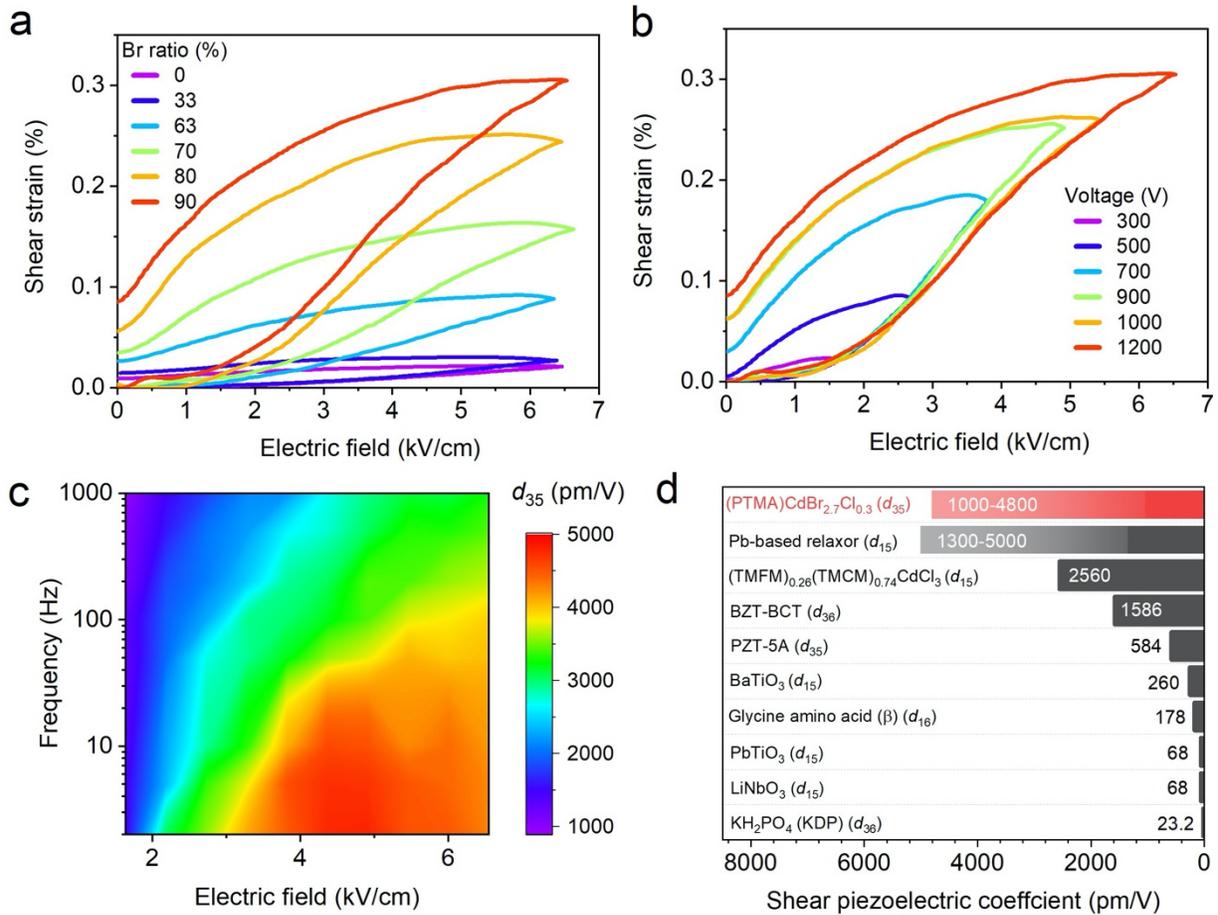

**Fig. 3 | Unipolar piezoelectric properties of (PTMA)CdBr$_{3x}$Cl$_{3(1-x)}$. a**, Unipolar *S-E* curves of selected compositions after pre-poling. **b**, Electric field dependent unipolar *S-E* curves of (PTMA)CdBr$_{2.7}$Cl$_{0.3}$. **c**, d$_{35}$ *versus* electric field at different frequencies. **d**, Comparison of shear piezoelectric coefficients of different materials (see Extended Data Table 3 for references).

To better understand the effect of Br substitution, we carried out *ab initio* density functional theory (DFT) calculations for (PTMA)CdCl$_3$ and (PTMA)CdBr$_3$. Figure 4a shows the minimum energy paths for switching between the two ferroelastic states, illustrating that the energy barrier in Br compound is greatly reduced with more flattened double-well landscape compared to the Cl counterpart. It is worth to note that the intermediate state of the ferroelastic switching is featured by centrosymmetric corner-sharing CdX$_4^-$ tetrahedra with the longest Cd-X bond being broken, a reminiscence of the HTP structure, while the switching process can be visualized as the bond switching (breaking and reformation) between Cd and the two adjacent apex halide atoms (see Supplementary Movies S3 and S4). We then evaluate the bond strength by calculating the crystal orbital Hamilton populations (COHPs) of the longest Cd-X bond, as well as the two shortest hydrogen bonds (X-H1 and X-H2) for comparison (Fig. 4b). Indeed, as characterized by the integrated COHPs up to the Fermi energy (Fig. 4b and Extended Data Table 4), stronger Cd-X and X-H bonds are found in the Cl compound, which is consistent with the higher electronegativity of Cl than Br, and the Cd-X bond is expected to have a dominant contribution to the overall barrier.

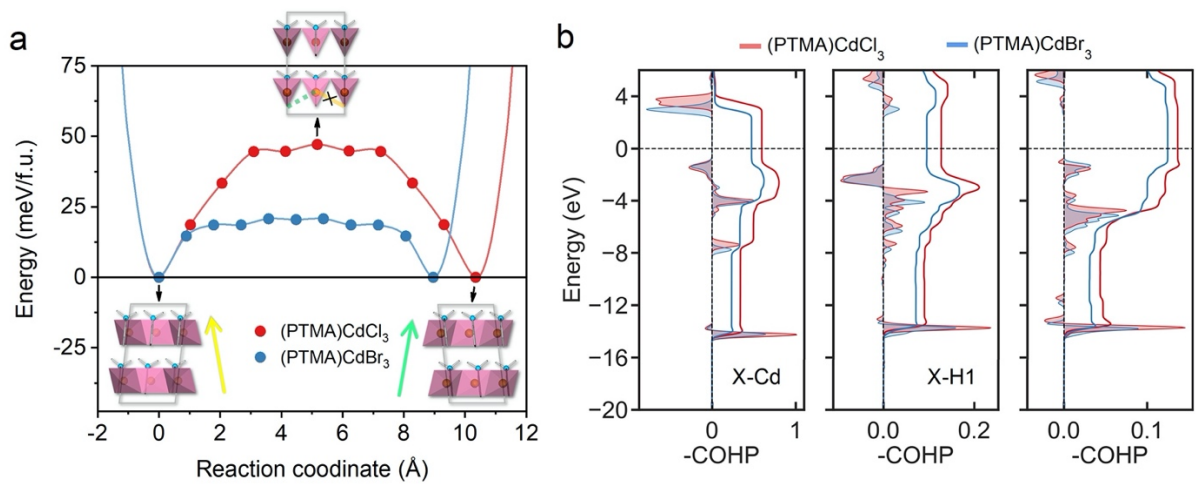

**Fig. 4 | Calculated switching barriers of (PTMA)CdBr$_3$ and (PTMA)CdCl$_3$. a**, Minimum energy paths calculated using the generalized solid-state nudged elastic band (GSSNEB) method. **b**, Calculated integrated crystal orbital Hamilton populations (ICOHPs) of the Cd-X bond that breaks during switching (left), and the two shortest hydrogen bonds (middle and right).

To summarize, we demonstrate that by geometric design of polarization switching pathways, it is possible to achieve colossal strain in a hybrid molecular ferroelectric, (PTMA)CdCl$_3$, through controlled ferroelastic transition. Furthermore, bond softening via chemical engineering is employed to soften the bonds and flatten the energy landscape for polarization rotation, leading to superior piezoelectric performance. The design strategy described here should inspire the development of next-generation piezoelectrics and electroactive materials based on OIHFs.

**Online Content** Methods, along with any additional Extended Data display items and Source Data, are available in the online version of the paper; references unique to these sections appear only in the online paper.

**METHODS**

**Synthesis of (PTMA)CdBr$_{3x}$Cl$_{3(1-x)}$ (0≤ x ≤1) single crystals.** All chemicals were purchased from Sigma-Aldrich and used as received. Single crystals of (PTMA)CdBr$_{3x}$Cl$_{3(1-x)}$ were synthesized by slow evaporation of filtered acetonitrile and deionized water mixture solutions (acetonitrile : deionized water=3:2 for 0≤ Br ratio ≤63%, 4:1 for 63%< Br ratio ≤100%) with stoichiometric ratios of PTMA-Cl, PTMA-Br, CdCl$_2$ and CdBr$_2$·4H$_2$O. Rod shape crystals with centimeter length were obtained upon slow evaporation of the solutions at room temperature for around three weeks. Pure Br crystal is pale blue. With higher Cl ratio, the color of the crystals becomes lighter. (PTMA)CdBr$_{3x}$Cl$_{3(1-x)}$ with Br ratios of 15%. 29%, 45%, 63%, 70%, 77%, 80% and 90% were synthesized by precursors with Br molar ratios of 20%, 33%, 50%, 67%, 73%, 80%, 87% and 96%, respectively. Br ratios of the crystals were calculated by C, H, and N mass fraction measured using an elemental analyzer (PerkinElmer model 2400 Series II).

**X-ray diffraction (XRD) measurement.** Powder XRD patterns were obtained by using Cu Kα radiation (λ=1.540598 Å, 40 kV and 30 mA) on a commercial diffractometer (Panalytical Xpert). Single crystal XRD patterns were collected using a Bruker APEX II diffractometer with Mo source (λ=0.71073 Å). Crystal structures were solved using Bruker SHELXTL Software Package and refined for all data by Full-matrix least-squares on F$^2$. All non-hydrogen atoms were subjected to anisotropic refinement. The hydrogen atoms were generated geometrically and allowed to ride in their respective parent atoms; they were assigned with appropriate isotropic thermal parameters and included in the structure factor calculations. The data can be obtained free of charge from the Cambridge Crystallography Data Center via www.ccdc.cam.ac.uk/data_request/cif

**DSC and TGA measurements.** Thermal analyses were performed using differential scanning calorimetry (DSC, TA INSTRUMENTS - Q10) by heating and cooling single crystal samples at a rate of 5 K min$^{-1}$ in an aluminum pan under nitrogen flow of atmospheric pressure. Thermogravimetric analyses (TGA, TA INSTRUMENTS – Q500) were conducted on a bulk single crystal with 77% Br at a rate of 10 K min$^{-1}$ in air.

**Dielectric and ferroelectric measurements.** Temperature dependent measurements were performed on a cryogenic micromanipulator probe station equipped with a heating stage. Crystals were carefully cut in the form of plates along *a*, *b* and *c* crystallographic axes, respectively. Silver paste was coated on top and bottom surfaces of the crystals as electrodes. Dielectric measurements were carried out using a commercial LCR meter (Agilent E4980A) with 1 V AC voltage. Ferroelectric properties were measured using a commercial ferroelectric tester (Precision LC, Radiant technologies) with high-voltage amplifier (Precision 4kV HVI, Radiant technologies).

**Piezoresponse force microscopy (PFM) measurements.** PFM was conducted using a commercial atomic force microscope (Asylum Research MFP-3D) with soft Au-coated silicon tips (Budget Sensor, ContGB-G, spring constant of ~0.2 N m$^{-1}$) to avoid sample surface damage. Resonance-enhanced dual AC resonance tracking (DART) mode (tip driving voltage=500mV, frequency ~220kHz) was used to enhance the signal-to-noise ratio.

**Shear strain and piezoelectricity measurements.** The measurement setup is shown in Extended Data Fig. 4. Crystals were firmly fixed on glass slide by epoxy with *a* axis perpendicular to the basal plane. Left and right sides of the crystals ((001) planes) were carefully cut and polished and silver paste was used as electrodes. An aluminum plate was attached on top of the crystal as the reflective mirror. Electric field induced movement of top surface (shear strain) was measured using a commercial photonic sensor (MTI-2100 Fotonic). All samples were pre-poled for piezoelectric measurements. Shear strain (S) is calculated by $S = d/H = \tan \Delta\beta$, where *d* is the displacement recorded by the photonic sensor, $H$ is the height of sample and $\Delta\beta$ is the shear angle. $d_{35}$ is calculated by $d_{35} = S_5/E_3$, where $E_3$ is electric field applied along $c$ axis.

**Computational methods.** (PTMA)CdCl$_3$ and (PTMA)CdBr$_3$ crystals were simulated with four formula units by density functional theory (DFT). The calculations were carried out using the projected augmented wave (PAW) method[28] and the Perdew-Burke-Ernzerhof (PBE) exchange-correlation functionals[29], as implemented in the Vienna ab-initio simulation package (VASP)[30]. The energy cutoff for the plane wave expansion was set to 500 eV, and a 2×2×4 Monkhorst-Pack k-mesh is used. We explicitly treated 12 valence electrons for Cd (4d$^{10}$5s$^2$), 4 for C (2s$^2$2p$^2$), 5 for N (2s$^2$2p$^3$), 1 for H (1s$^1$), 7 for Cl (3s$^2$3p$^5$) and 7 for Br (4s$^2$4p$^5$). The initial structures were obtained from the experimental crystallographic data, then fully optimized until all atomic forces were less than 0.002 eV/Å. The effects of van der Waals (vdW) interactions[31] were included during the structural and electronic relaxations, as well as all the subsequent calculations. The ferroelectric polarizations were calculated by the Berry phase approach[32,33], and the elastic constants were determined from the strain-stress relationship. The minimum energy paths for polarization switching were calculated using the generalized solid-state nudged elastic band (GSSNEB) method[34]. We used the LOBSTER package to compute the crystal orbital hamiltonian populations (COHP)[35-37], with a 3×3×5 Monkhorst-Pack k-mesh.

**Acknowledgements:** We thank P. J. S. Buenconsejo (Facility for Analysis, Characterisation, Testing and Simulation, Nanyang Technological University) for assistance on XRD measurements, and X. R. Zhou (School of Materials Science and Engineering, Nanyang Technological University) for discussion on piezoelectric measurements. L.Y. and B.X. acknowledge the startup funds from Soochow University, and the support from Priority Academic Program Development (PAPD) of Jiangsu Higher Education Institutions. L.Y. also acknowledges the support from the National Natural Science Foundation of China (11774249), the Natural Science Foundation of Jiangsu Province (BK20171209), and the Key University Science Research Project of Jiangsu Province (18KJA140004). H.J.F. acknowledges the support from AME Individual Research Grant (Grant number: A1883c0004), Agency for Science, Technology, and Research (A*STAR). J. W. acknowledges the support from the Ministry of Education, Singapore (Grant numbers: AcRF Tier 1 118/17 and 189/18) and the startup grant from Southern University of Science and Technology (SUSTech), China.

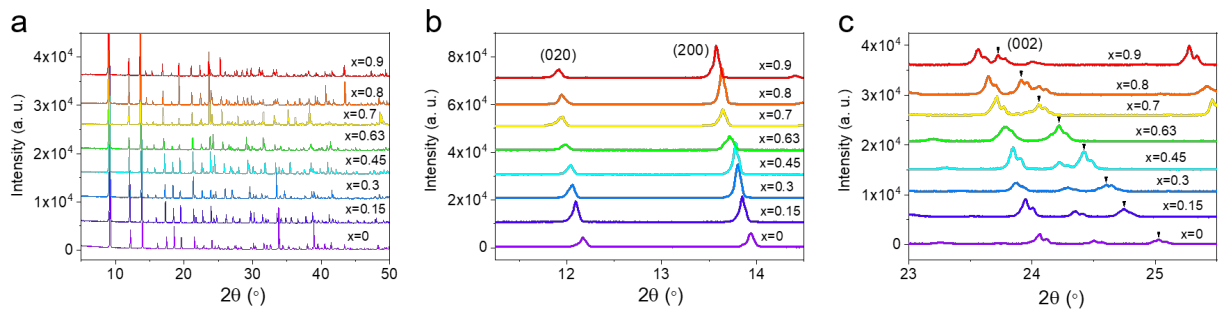

**Extended Data Fig. 1 | Powder XRD patterns of (PTMA)CdBr$_{3x}$Cl$_{3(1-x)}$ with different Br contents at room temperature.** θ−2θ scan from **a**, 5° to 50°, **b**, around (020) and (200) peaks, and **c**, around (002) peaks.

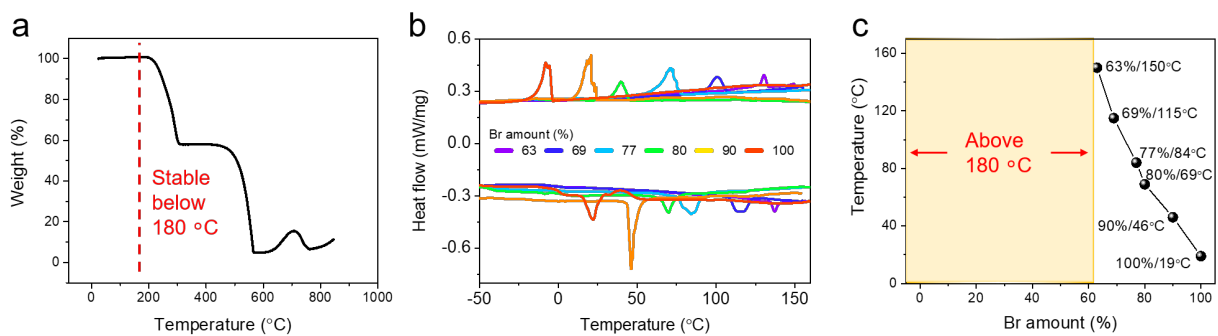

**Extended Data Fig. 2 | Thermal properties of (PTMA)CdBr$_{3x}$Cl$_{3(1-x)}$. a**, TGA analysis of a 77%-Br crystal. **b**, DSC curves of heating-cooling cycles for Br-rich samples. **c**, Phase transition temperatures for samples with different Br contents.

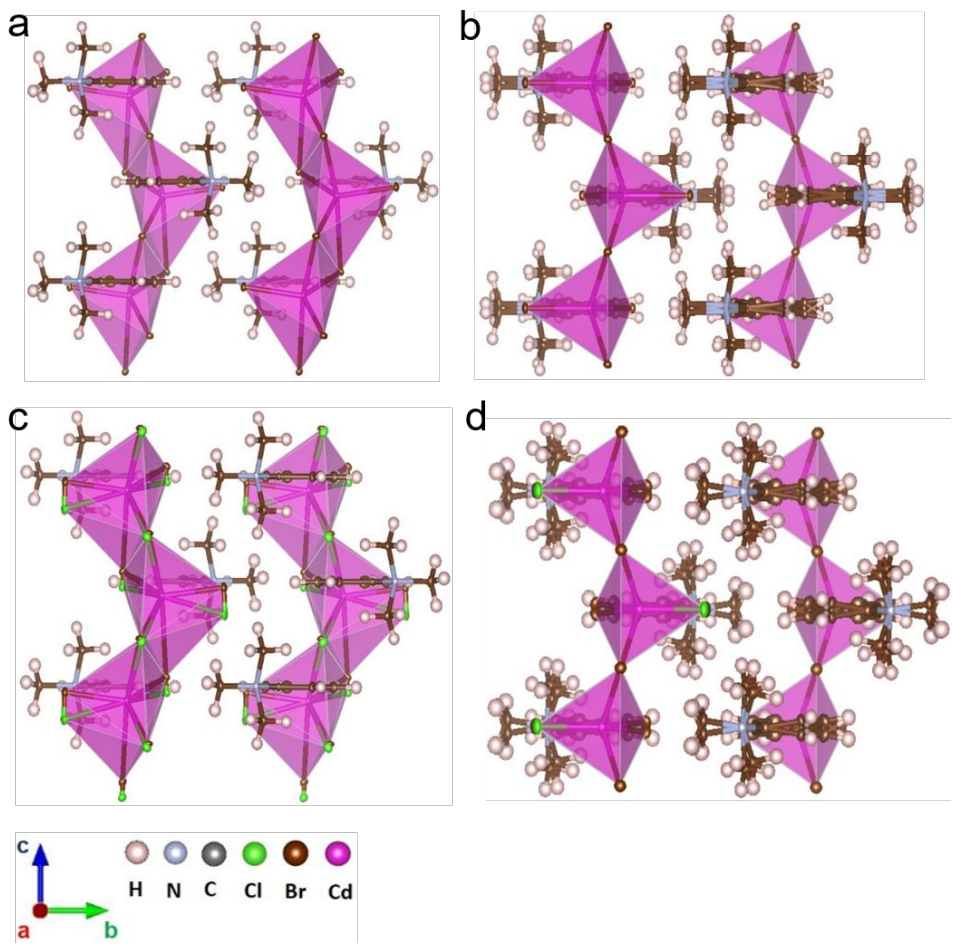

**Extended Data Fig. 3 | Crystal structures of Br rich samples.** **a**, **b**, Structures of pure Br compound at low and high temperature, respectively. **c**, **d**, Structures of 76% Br compound at low and high temperature, respectively.

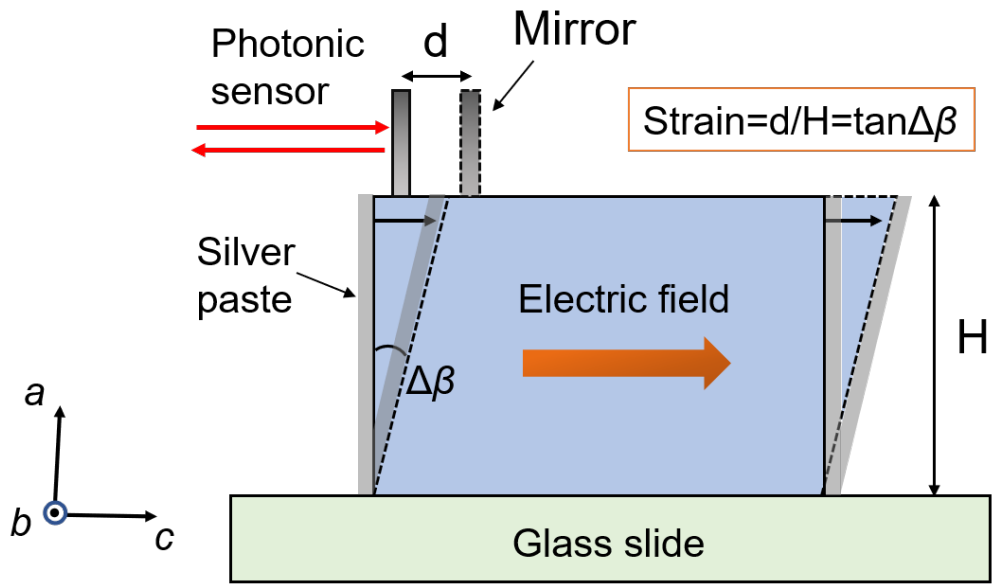

**Extended Data Fig. 4 | Experimental setup for the shear strain and piezoelectricity measurements.** The coordinate indicates the crystallographic orientation of the measured sample.

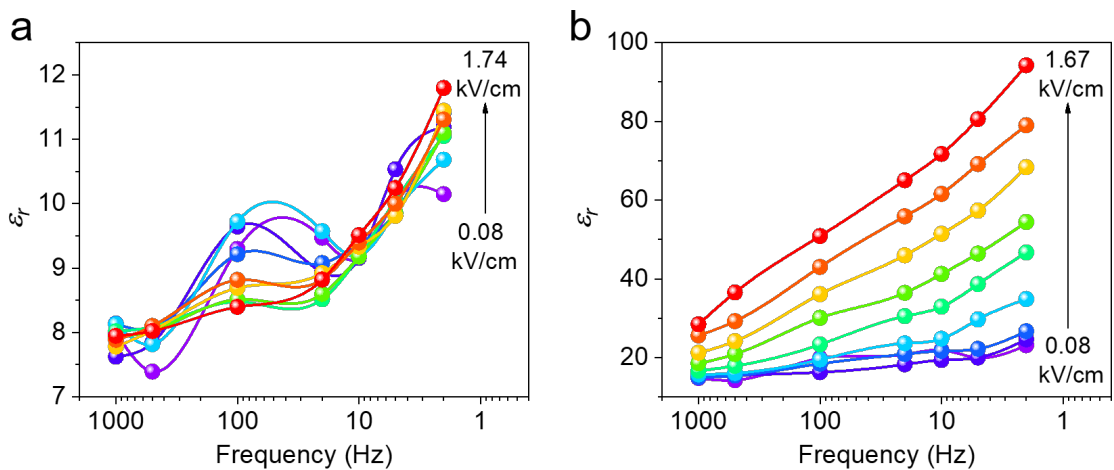

**Extended Data Fig. 5 | Electric field and frequency dependent dielectric responses** of **a**, pure Cl and **b**, 90%-Br samples.

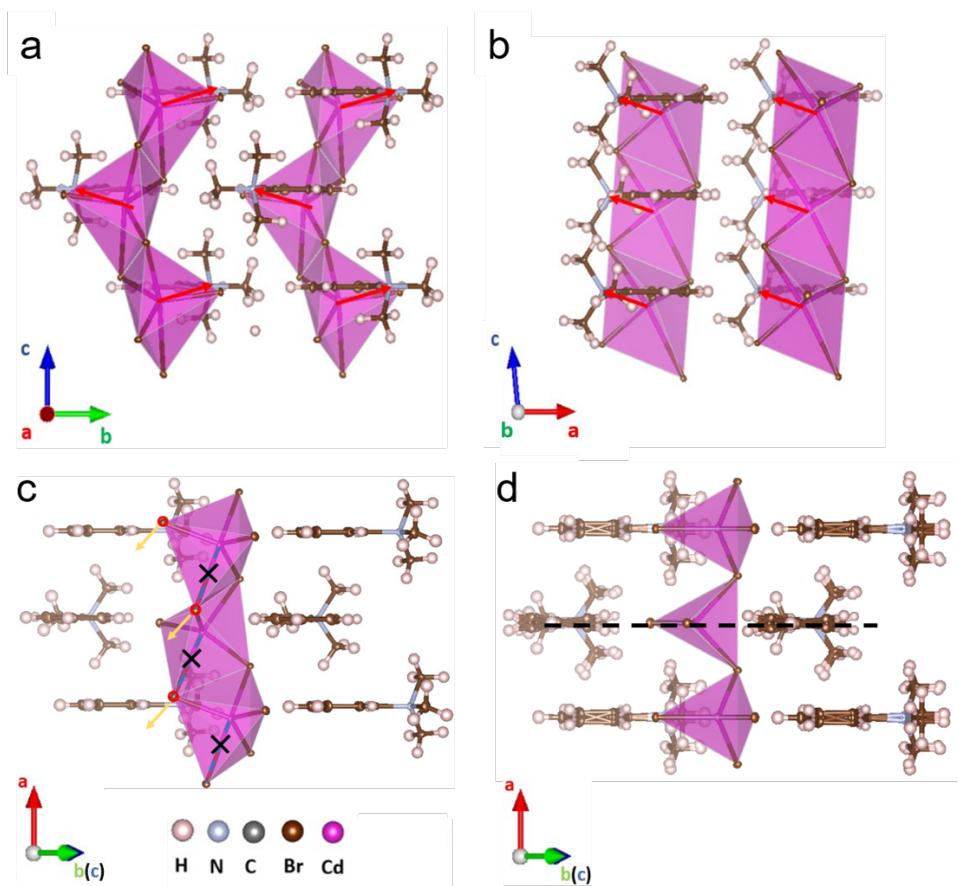

**Extended Data Fig. 6 | Polarization and phase transition analysis of (PTMA)CdBr$_{3x}$Cl$_{3(1-x)}$, using pure Br as example. a**, **b**, Red arrows indicate dipoles of unit cell. At low temperature phase, polarization along *c* and *a* axis can be identified from structure perspectives in *a* and *b* directions. **c**, Structure change at phase transition temperature. Broken Cd-Br bonds are labeled by blue dash lines and black crossings. Orange arrows indicates the corresponding movement of Br atom in red cycles. **d**, High temperature phase structure. Black dash line labels the induced mirror symmetry.

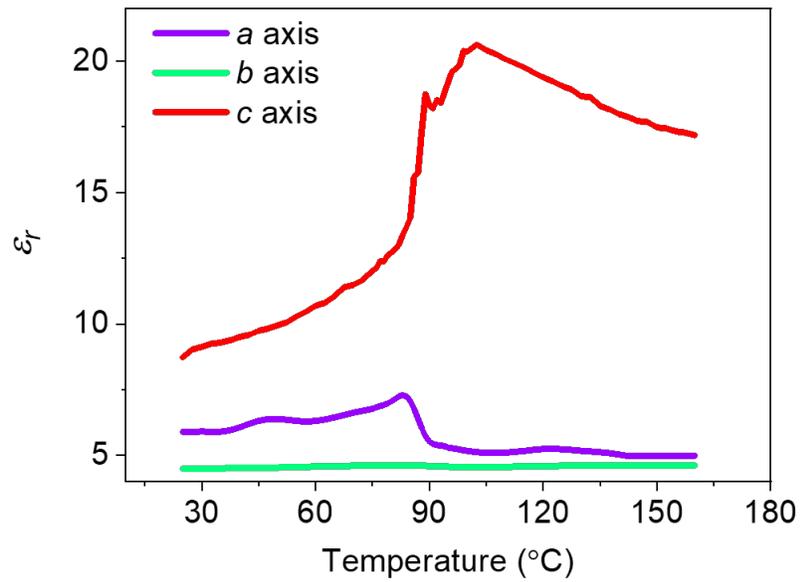

**Extended Data Fig. 7 | Temperature dependent dielectric response** of a 77%-Br sample along different crystallographic axes in heating process with a frequency of 1 kHz.

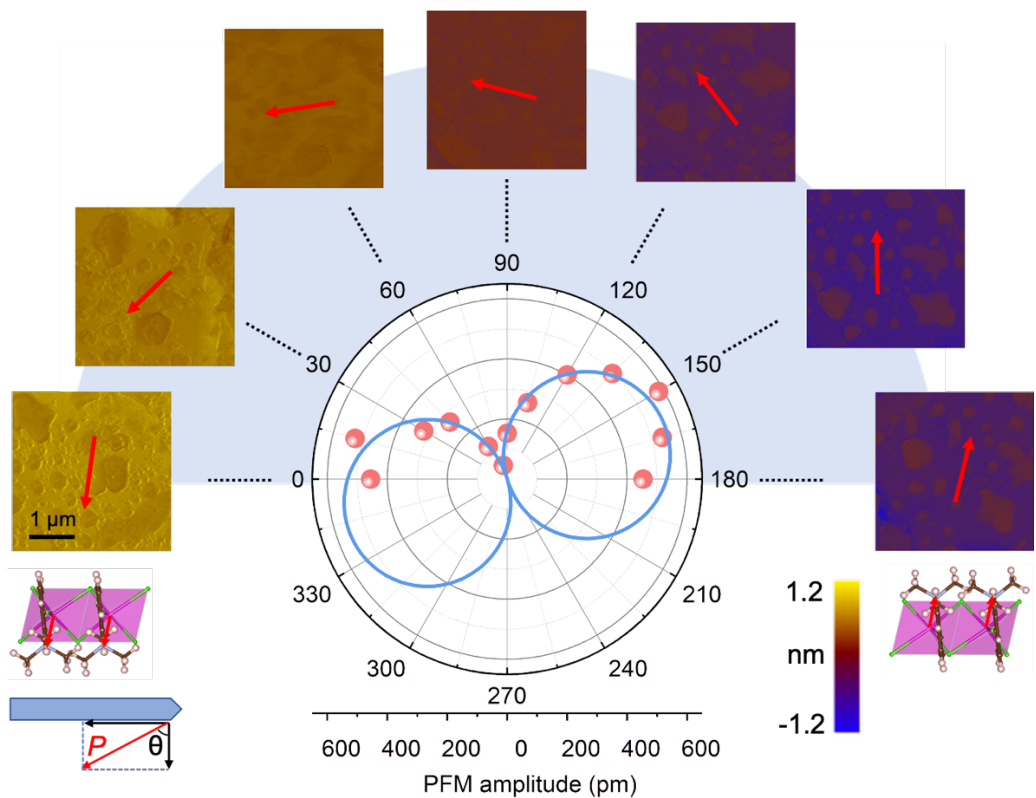

**Extended Data Fig. 8 | Angular PFM study of a 77%-Br single crystal.** PFM images are collected with the crystal oriented at different azimuthal angles with regard to the cantilever axis. The average PFM amplitude of each image is plotted in polar coordinate. Initial and final crystallographic orientations are indicated beneath the PFM images, with red arrow denotes the polarization vector.

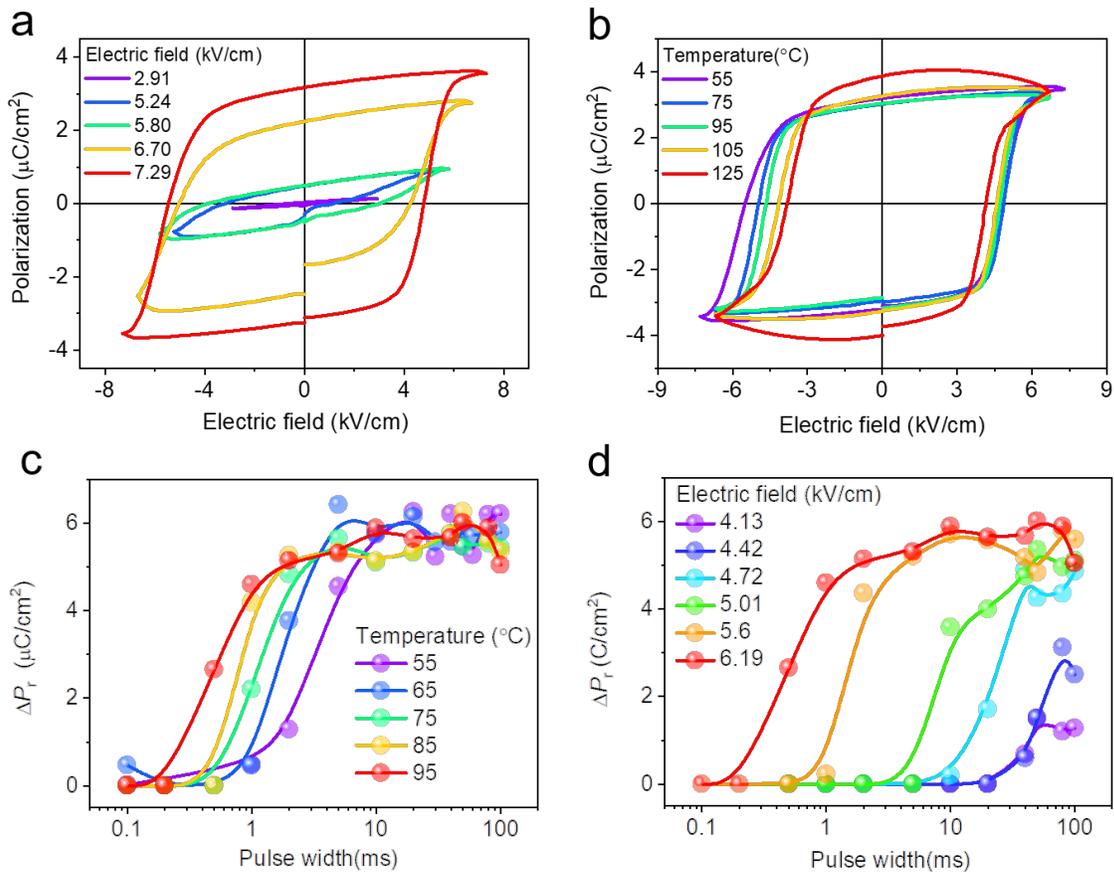

**Extended Data Fig. 9 | Detailed ferroelectric characterizations of a 63%-Br sample.** The frequency of P - E loop measurements is 2 Hz. a, Hysteresis loops under different electric fields at 55 °C. b, Temperature dependent P - E loops. c, PUND measurements at different temperatures under an electric field of 6.19 kV/cm. d, Electric field dependent PUND measurements at 95 °C.

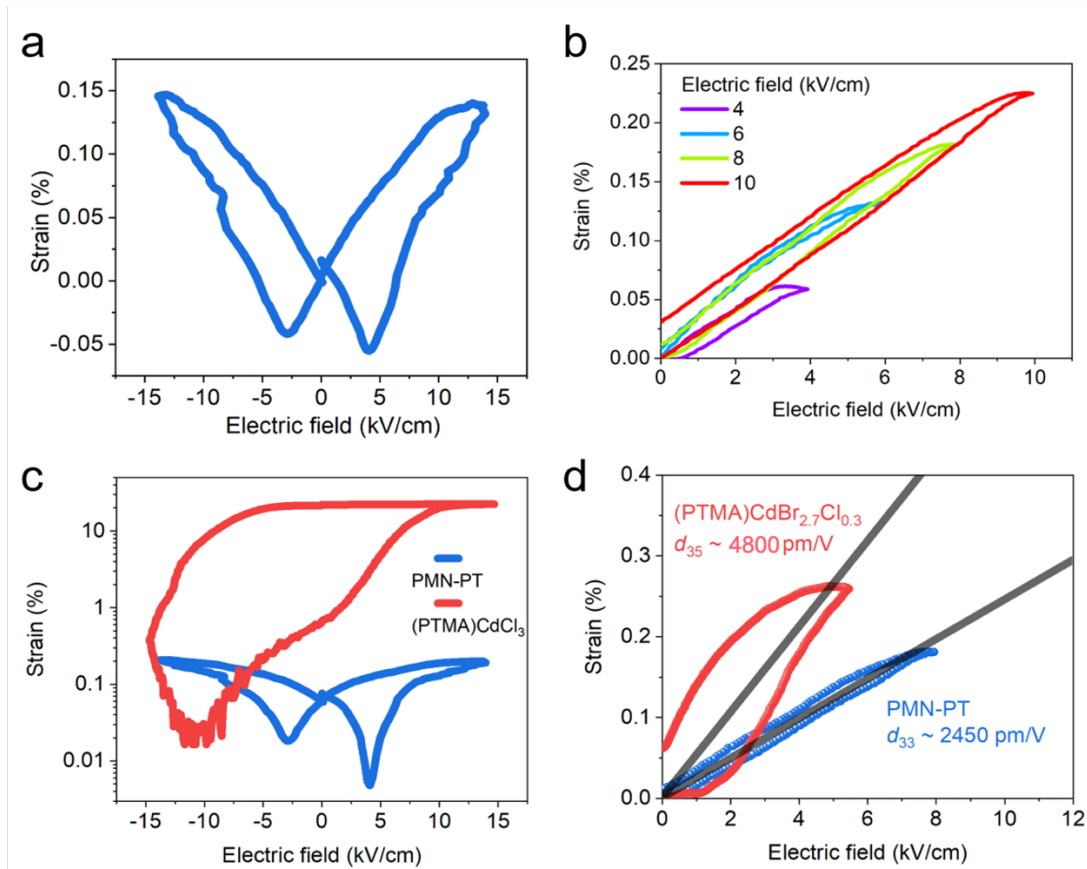

**Extended Data Fig. 10 | Comparison of the electrostrain between PMN-PT and (PTMA)CdBr$_{2.7}$Cl$_{0.3}$.** a, Bipolar S – E butterfly loop of PMN-PT. b, Unipolar piezoelectric response of PMN-PT under different electric fields. c, Comparison of the switchable electrostrains in logarithm scale. d, Comparison of the piezoelectric responses. The slopes of the linear fits are the piezoelectric coefficients.

**Extended Data Table 1.** Crystallographic data of $(PTMA)CdBr_{3X}Cl_{3(1-X)}$.

| Br amount (%) | 0 | 76 | | 100 | |
|---|---|---|---|---|---|
| Empirical formula | $(PTMA)CdCl_3$ | $(PTMA)CdBr_{2.3}Cl_{0.7}$ | | $(PTMA)CdBr_3$ | |
| Temperature (K) | 299(2) | 296(2) | 393(2) | 253(2) | 308(2) |
| Space group | Cc | Cc | Ama2 | Cc | Ama2 |
| a (Å) | 12.7404(4) | 12.9705(15) | 12.9801(13) | 13.0388(5) | 13.0309(9) |
| b (Å) | 14.4928(6) | 14.7252(17) | 14.7100(14) | 14.7738(6) | 14.7328(9) |
| c (Å) | 7.1432(3) | 7.4093(8) | 7.6583(6) | 7.4720(3) | 7.7137(4) |
| β (°) | 96.3055(14) | 95.121(4) | 90 | 95.2538(10) | 90 |
| Volume (Å³) | 1310.97(9) | 1409.48 | 1462.25 | 1433.3 | 1480.89 |
| Z | 4 | 4 | 4 | 4 | 4 |
| F (000) | 696 | 861 | 859 | 912 | 912 |
| R factor (%) | 3.29 | 3.8 | 3.15 | 3.71 | 3,88 |
| Goodness-of-fit | 1.029 | 1.036 | 1.009 | 1.072 | 1.064 |

**Extended Data Table 2**. Actuation properties of different material systems

| Material system | Compound | Strain (%) | Stress (MPa) | Strain type | Work per volume J/m$^3$ | Power per volume W/m$^3$ | Frequency (Hz) | Reference |
|---|---|---|---|---|---|---|---|---|
| Hybrid ferroelectric | (PTMA)CdCl$_3$ | 21.5 | 25 | Shear | 3.2×10$^6$ | 4×10$^6$ | 2 | This work |
| Ferro/Piezoelectric oxides | Soft PZT | 0.1 | 20 | Shear | 2×10$^4$ | 2×10$^5$ | 10 | 1 |
| | PMN-0.3PT | 0.4 | 35 | Normal | 7×10$^4$ | 7×10$^5$ | 10 | 2 |
| | BNT-BT-KNN | 0.22 | 120 | Normal | 1.3×10$^5$ | 6.5×10$^6$ | 50 | 3 |
| | PMN-0.38PT | 0.66 | 20 | Normal | 6.6×10$^4$ | 6.6×10$^6$ | 100 | 4 |
| | BTO | 0.93 | 6 | Normal | 3.7×10$^4$ | 3.7×10$^7$ | 1000 | 4 |
| Shape memory alloy | NiTi | 8 | 700 | Normal | 2.8×10$^7$ | | | 5-7 |
| | NiTiHf | 3.5 | 150 | Shear | 5.25×10$^6$ | | | 8 |
| | CuZnAl | 4 | 550 | Normal | 1×10$^7$ | | | 5-7 |
| | CuAlNi | 4 | 450 | Normal | 9×10$^6$ | | | 5-7 |
| MSMAs | NiMnGaCoCu | 12 | 2 | Normal | 1.7×10$^5$ | | | 9 |
| | 14M NiMnGa | 9.5 | 1.8 | Normal | 1.45×10$^5$ | | | 10 |
| | NiFeGaCo | 8.5 | 8.5 | Normal | 7.225×10$^5$ | | | 11 |
| Shape memory polymer | Polyurethane | 80 | 0.05 | Normal | 2×10$^4$ | 20 | 1×10$^{-3}$ | 12 |
| | Styrene resin | 30 | 0.236 | Shear | 3.5×10$^4$ | 35 | 1×10$^{-3}$ | 13 |
| | Semicrystalline oligome | 50 | 0.5 | Shear | 1.2×10$^5$ | 120 | 1×10$^{-3}$ | 14 |
| | Nylon 66 | 10 | 22 | Normal | 1×10$^6$ | 5×10$^6$ | 1-5 | 15 |
| Electroactive polymers | Graft elastomer | 4 | 22 | Normal | 4.4×10$^5$ | 4.4×10$^7$ | 100 | 16 |
| | Polyurethane | 11 | 2 | Normal | 1.1×10$^5$ | 1.1×10$^7$ | 100 | 16 |

| | Material | | | | | | | |
|---|---|---|---|---|---|---|---|---|
| | VHB 4910 acrylic | 68 | 2.4 | Normal | $1.36\times10^6$ | $5.4\times10^7$ | 40 | 17 |
| | HS3 silicone | 54 | 0.4 | Normal | $1.6\times10^5$ | $3.2\times10^8$ | 2000 | 17 |
| | CF19-2186 | 39 | 0.8 | Normal | $2\times10^5$ | $3.4\times10^7$ | 170 | 17 |
| | Liquid crystal | 4 | 0.12 | Normal | $2.4\times10^3$ | $2.4\times10^5$ | 100 | 18 |
| | Polypyrrole | 5 | 1.2 | Normal | $3\times10^4$ | $3\times10^2$ | 0.01 | 19 |
| Piezoelectric polymer | PVDF | 1.6 | 35 | Normal | $2.8\times10^5$ | $5.6\times10^5$ | 2 | 20 |
| | PVDF+CuPc | 2 | 15 | Normal | $1.5\times10^5$ | $1.5\times10^5$ | 1 | 21 |
| | Irradiated PVDF | 4 | 15 | Normal | $3\times10^5$ | $3\times10^6$ | 1-10 | 22 |

**Table S3.** Shear piezoelectric properties of different materials.

| Compound | Piezoelectric coefficient (pm/V) | Piezoelectric mode | Reference |
|---|---|---|---|
| (PTMA)CdBr$_{2.7}$Cl$_{0.3}$ | 1000-4800 | $d_{35}$ | This work |
| KH$_2$PO$_4$ | 23.2 | $d_{36}$ | 1 |
| LiNbO$_3$ | 68 | $d_{15}$ | 1 |
| PbTiO3 | 68 | $d_{15}$ | 2 |
| Glycine amino acid (β) | 178 | $d_{16}$ | 3 |
| BaTiO$_3$ | 260 | $d_{15}$ | 1 |
| PZT-5A | 584 | $d_{35}$ | 4 |
| BZT-BCT | 1586 | $d_{36}$ | 5 |
| (TMFM)$_{0.26}$(TMCM)$_{0.74}$CdCl$_3$ | 2560 | $d_{15}$ | 6 |
| Pb Based relaxor | 1300-5000 | $d_{15}$ | 7 |

**Extended Data Table 4.** Calculated bond lengths and integrated crystal orbital Hamilton populations (ICOHP) up to the Fermi energy of the X-Cd bond that breaks during switching, the shortest (X-H1) and the second shortest hydrogen bond for pure Cl and pure Br structures.

| Bond | Bond length (Å) | ICOHP (eV) |
| --- | --- | --- |
| Cd-Cl | 2.81760 | -0.59455 |
| Cl-H1 | 2.61135 | -0.12741 |
| Cl-H2 | 2.71702 | -0.13675 |
| Cd-Br | 3.02990 | -0.47219 |
| Br-H1 | 2.69734 | -0.09563 |
| Br-H2 | 2.84170 | -0.12472 |